# 2D titanium carbonitrides and their hydroxylated derivatives: Structural, electronic properties and stability of *MX*enes $Ti_3C_{2-x}N_x$ and $Ti_3C_{2-x}N_x(OH)_2$ from SSC-DFTB calculations

A. N. Enyashin,* A.L. Ivanovskii

*Institute of Solid State Chemistry, Ural Branch of the Russian Academy of Sciences, 620041 Ekaterinburg, Russia*

**Abstract**

3D titanium carbonitrides $TiC_xN_y$ possess excellent physical, chemical, and mechanical properties, attractive for various industrial applications. Most recently, the uncommon nano-sized layers of 2D-like titanium carbonitrides were fabricated from MAX phases. Herein, the structural, electronic properties and stability of these new compounds as well as their hydroxylated derivatives – so-called *MX*enes $Ti_3C_{2-x}N_x$ and $Ti_3C_{2-x}N_x(OH)_2$ are probed by means of SSC-DFTB calculations. The genesis of the properties is discussed in the sequence: binary *MX*enes $Ti_3C_2$ ($Ti_3N_2$) → their hydroxylated forms $Ti_3C_2(OH)_2$ ($Ti_3N_2(OH)_2$) → pristine *MX*ene $Ti_3C_{2-x}N_x$ → hydroxylated $Ti_3C_{2-x}N_x(OH)_2$. Our results show that the examined materials are metallic-like. The most favorable type of OH-covering is presented by the occupation of the hollow sites between three neighboring carbon (nitrogen) atoms. The formation of 2D *MX*ene carbonitrides with random distribution of C and N atoms was found to be thermodynamically more favorable.

## 1. Introduction

Titanium monocarbide (TiC) and mononitride (TiN) have attracted considerable interest because of unique combination of properties such as high melting temperature, high hardness, high electrical and thermal conductivities, good wear, radiation and corrosion resistance, and are widely used for cutting tools, wear-resistant coatings, and for other technological applications [1-5]. These cubic (NaCl-like) phases possess complete mutual solubility, forming a continuous series of solid solutions - titanium carbonitrides, $TiC_xN_{1-x}$, which, in turn, show a set of superior properties, incorporating the advantages of both TiC and TiN, and being attractive for industrial materials such as cermets. Recently, for further improving of the performance of these materials, the considerable efforts have been undertaken for the synthesis of submicron and nano-sized titanium carbonitrides [6-8].

Most recently, Naguib et al. [9] have proposed an elegant exfoliation approach to prepare a new family of 2D-like nano-sized binary and ternary transition metal carbides and nitrides (which are known today also as *MXenes*) from MAX phases, providing another promising stage in the design of the materials with intriguing functionalities and applications.

The bulk MAX phases (with the general formula $M_{n+1}AX_n$, where n = 1, 2, or 3; M are transition *d* metals from groups III-VI; X are C or N; and A are *sp* elements) consist of stacked carbide (nitride) blocks [$M_{n+1}X_n$] and planar A atomic sheets (reviews [10-14]). The system of inter-atomic interactions in MAX phases is very anisotropic with strong directional M-X bonds within the blocks [$M_{n+1}X_n$], whereas the bonds between A atomic sheets and blocks [$M_{n+1}X_n$] are relatively weak. Therefore, the basic idea [9] was to remove the most weakly bonded sheets of A atoms from the MAX phases and to obtain free-standing carbide (nitride) blocks [$M_{n+1}X_n$] as novel 2D-like materials.

Successful realization of this idea has allowed today to synthesize a quite broad group of binary *MX*enes: $Ti_2C$, $Ti_3C_2$, $Ta_4C_3$ and some others [9,15], for which a set of physical and chemical properties were studied experimentally [9,15-17] or predicted theoretically [9,18-22]. Moreover, by the immersion of the powder of MAX phase $Ti_3AlCN$ at room temperature in HF followed by sonication, the more complex *MXene* $Ti_3CN$, *i.e.* nano-sized 2D titanium carbonitride, was recently prepared [15].

In this paper we focus on novel 2D titanium carbonitride to give an insight into the structural and electronic properties of this uncommon material. Noteworthy, that already in the first experiments [9] it has been revealed, that the fabricated *MXenes* are not pristine carbide (nitride) nano-layers, but are terminated by various functional groups. So, after immersion of $Ti_3AlC_2$ in hydrofluoric acid, the reaction $Ti_3AlC_2 + 3HF = AlF_3 + 3/2H_2 + Ti_3C_2$, which allows to obtain the required 2D $Ti_3C_2$, is accompanied by the reactions $Ti_3C_2 + 2H_2O = Ti_3C_2(OH)_2 + H_2$ and $Ti_3C_2 + 2HF = Ti_3C_2F_2 + H_2$ leading to the formation of hydroxylated or fluorinated $Ti_3C_2$ derivatives [9]. The presence of surface functional groups leads to drastic changes of properties of *MXenes* [9,18-22]. Therefore, the surface chemistry may open a wide range of possibilities to design a variety of new 2D materials with variable properties by modulating the type and degree of termination of *MXenes* by various adatoms or molecules.

For the discussed *MXene* $Ti_3CN$, the available data of energy dispersive X-ray spectroscopy testify [15] that the synthesized samples also are terminated with OH and/or F surface groups. Therefore, along with pristine 2D titanium carbonitride, its hydroxylated derivatives will be studied.

## 2. Models and computational aspects

The calculations of aforementioned 2D titanium carbonitrides were launched by the consideration of binary *MX*enes $Ti_3C_2$ and $Ti_3N_2$. Their atomic models were constructed by removing of Al sheets from the parent 312 MAX phases $Ti_3AlC_2$ and $Ti_3AlN_2$. Then, simulating free-standing pristine *MXenes* the adjacent blocks [$Ti_3C_2$] ([$Ti_3N_2$]) were separated from each other by a vacuum region to exclude inter-block interactions. These 2D materials ($Ti_3C_2$ or $Ti_3N_2$) consist of five atomic sheets with a hexagonal-like unit cell, where two carbon (or nitrogen) atomic sheets are sandwiched between the three Ti sheets: [Ti2-C-Ti1-C-Ti2] ([Ti2-N-Ti1-N-Ti2]), with the atoms placed in positions Ti1: 2*a* (0, 0, 0,); Ti2: 4*f* (⅔, ⅓, *v*); and C(N): 4*f* (⅓, ⅔, *u*), see Fig. 1.

Afterwards, the atomic models of hydroxylated binary *MX*enes were designed, assuming their full surface termination by hydroxy groups: $Ti_3C_2(OH)_2$, and $Ti_3N_2(OH)_2$. Here, two main configurations of OH covering of external Ti2 sheet are admissible (see also [18,21]), where all OH groups are placed at the hollow sites between three neighboring carbon (nitrogen) atoms (position A), or these groups are placed at the top site of the carbon (nitrogen) atoms (position B), Fig. 1. Accordingly, for the double-sided surface termination of *MX*enes, three main configurations occur: two symmetric (AA and BB), and asymmetric configuration AB, Fig. 1.

Furthermore, the more complex mixed-anion *MX*enes with the formal stoichiometry $Ti_3CN$ synthesized in the experiments [15] were examined. Here, two main types of mutual C/N atomic distributions are possible: either ordered configuration (**I**): ([Ti2-C-Ti1-N-Ti2]), where carbon and nitrogen atoms form the individual atomic sheets, or numerous configurations of the type [Ti2-($C_{1-x}N_x$)-Ti1-($N_{1-x}C_x$)-Ti2] (**II**) with C and N atoms distributed in adjacent sheets, see Fig. 2.

In turn, a much complex situation arises for *MXenes* $Ti_3CN$ (**I, II**) with surface functionalization by OH groups, where numerous types of configurations emerge depending both from the mutual C/N atomic distributions within *MXene* (**I, II**), and from a type of surface termination by OH groups. In our study, we have chosen some characteristic configurations for $Ti_3CN(OH)_2$, which are depicted in Fig. 2. These configurations combine the aforementioned types (**I** *versus* **II** plus A *versus* B) and allow us to analyze the main trends in stability, structural and electronic properties of examined $Ti_3CN(OH)_2$ as depending from C/N distributions and from the type of surface termination.

All calculations of structural and electronic properties of pure and functionalized MXene layers were performed within the self-consistent charge density-functional tight-binding method (DFTB) [23,24]. This quantum-mechanical method was earlier applied for the simulations of various titanate nanostructures and organic adsorbates on titanium oxide [25-27]. It has been found to describe these systems in reasonable agreement with experimental data and high-level theoretical methods at lower computational cost. The unit cells of all structures considered in this work were fully relaxed preserving hexagonal symmetry under periodic boundary conditions as implemented in Dylax software using the conjugate gradient algorithm and 10 *k*-points for each direction of two-dimensional reciprocal lattice. The supercells of $Ti_3C_{2-x}N_x$ and $Ti_3C_{2-x}N_x(OH)_2$ layers have included 5×5 unit cells and have been relaxed under periodic boundary conditions in Γ-point approximation. The self-consistent calculations were considered to be converged when the difference in the total energy did not exceed $10^{-3}$ eV/atom as calculated at consecutive steps and the forces between atoms are close to zero.

## 3. Results and discussion.

### *3.1. $Ti_3C_2$, $Ti_3N_2$, $Ti_3C_2(OH)_2$, and $Ti_3N_2(OH)_2$.*

We start our study with the discussion of reference systems: pristine 2D *MXenes* $Ti_3C_2$, and $Ti_3N_2$, - as free-standing carbide (nitride) blocks, extracted from corresponding 3D MAX phases $Ti_3AlC_2$ and $Ti_3AlN_2$. Upon the full geometry optimization, the relaxed structures of these 2D materials preserve the integrity of starting geometry. In agreement with the previous data [9,18-21], the results reveal that both $Ti_3C_2$, and $Ti_3N_2$ are metallic-like compounds (Fig. 3) with high density of states at the Fermi level ($E_F$, Fig. 4), leading to the magnetic instability of these materials, see [20].

From the total and partial densities of states (DOSs, Fig. 4) we see that the valence states of *MXene* $Ti_3C_2$ are divided into two main subbands. The subband centered at about -3 eV below the Fermi level is composed by hybridized states C 2*p* -

Ti 3*d*, which form the covalent Ti-C bonds. The near-Fermi subband contains mainly Ti 3*d* states, which are responsible for metallic-like conductivity of this material. The valence spectrum of *MXene* $Ti_3N_2$ also contains two main subbands, but possesses some characteristic features. The subband of the hybridized states N2*p*-Ti3*d*, which are responsible for covalent Ti-N bonds, becomes narrower and is shifted down at about 2 eV. On the contrary, the near-Fermi Ti 3*d* –like subband becomes wider. Thus, the gap between these subbands is increased. These results agree well with the earlier data [16] as obtained within *ab initio* projector augmented waves (PAW) approach, VASP code.

The structural parameters and relative stability of hydroxylated *MX*enes $Ti_3C_2(OH)_2$, and $Ti_3N_2(OH)_2$ are compared as depending from the examined configurations of OH covering of external Ti sheets om Fig. 5 and Table 1. It can be noticed that for these hydroxylated derivatives (i) their lattice parameters depend strongly on the type of OH covering, (ii) the parameters for all of $Ti_3N_2(OH)_2$ configurations are smaller, than for $Ti_3C_2(OH)_2$, and (iii) thetrends in the change of the lattice parameters for $Ti_3C_2(OH)_2$, and $Ti_3N_2(OH)_2$ with the same type of covering are similar.

The understanding of stability of different *MX*enes depending on the surface arrangement of hydroxy groups can be elucidated from the relative total energies $E_{tot}$ listed in Table. These data reveal that for both $Ti_3C_2(OH)_2$, and $Ti_3N_2(OH)_2$, the location of OH groups at the positions A, *i.e.* above the hollow sites between the three neighboring carbon (nitrogen) atoms, is the most favorable. The hydroxylated *MXenes* with the double-sided AA type of the surface termination are the most stable. On the contrary, the position B (above the Ti atoms) is found as the least stable. Thus, the intermediate type of the double-sided covering (AB configuration) presents the intermediate case. Similar results are found for hydroxylated forms of lower *MXenes* homologues with compositions $M_2$C, and $M_2$N, where $M$ = Ti [18,21], Sc, V, Cr, Nb, and Ta [21].

The most stable surface modified *MXenes* retain metallic-like behavior like their parent phases, Fig. 3. The most evident changes in electronic spectra of $Ti_3C_2$, and $Ti_3N_2$ after their hydroxylation include the presence of additional O2*p*-like subbands. For $Ti_3C_2(OH)_2$ this subband is located below the subband of hybridized C2*p*-Ti3*d* states, whereas for $Ti_3C_2(OH)_2$ this O2*p* - and N2*p*-Ti3*d*-like subbands partially merge,

Fig. 4. Besides, the formation of directional *pd* bonds between oxygen of OH groups and the titanium atoms of external Ti2 sheet leads to partial depopulation of the near-Fermi states. A decreasing of DOSs at the Fermi level should offend against the magnetic states of pristine *MXenes* [18], determining the non-magnetic behavior of their hydroxylated derivatives.

### *3.2. $Ti_3CN$, and $Ti_3CN(OH)_2$*

In this section the 2D titanium carbonitride $Ti_3CN$ and its hydroxylated derivatives are discussed, assuming that carbon and nitrogen atoms form the individual atomic sheets, *i.e.* the ordered configuration **I**, Fig. 2.

The partial replacement of C on N atom in the unit cell of MXene layer leads to formation of non-equivalent sheets Ti2-C-Ti1 and Ti2-N-Ti1 within single layer. After geometry optimization an unilateral contraction of a layer [Ti2-C-Ti1-N-Ti2] can be found from the side of nitrogen-containing sheet. Indeed, the estimated ratio of the distances $d^{Ti\text{-}C\text{-}Ti}/d^{Ti\text{-}N\text{-}Ti} \sim 1.65$. The electronic bands intersect $E_F$, revealing the metallic-like conductivity of $Ti_3CN$, Fig. 6. In the electronic spectrum of $Ti_3CN$ three separate subbands are presented, which are composed by hybridized N2*p*-Ti3*d*, hybridized C2*p*-Ti3*d*, and near-Fermi Ti3*d* states, Fig. 7.

For hydroxylated derivatives of $Ti_3CN$, taking into account the non-equivalent surrounding of two external Ti2 sheets (by carbon or nitrogen atoms) already four types of the double-sided surface termination arise: AA, AB, BA and BB, Fig. 2. The estimations of their relative total energies demonstrate that similar to $Ti_3C_2(OH)_2$, and $Ti_3N_2(OH)_2$ the most favorable arrangement of hydroxyl groups is AA type (Table 1). Noteworthy, the most dominanting factor determing the stability would be the surface arrangement of the external Ti2 sheet located over nitrogen sheet, since $(E_{tot}^{AB} - E_{tot}^{AA}) > (E_{tot}^{BA} - E_{tot}^{AA})$, see Table 1.

The lattice parameters of $Ti_3CN(OH)_2$, are close to the averaged values of the parameters corresponding to $Ti_3C_2(OH)_2$, and $Ti_3N_2(OH)_2$. As well, there is the same trend in their change depending on the type of covering. The one exception is the lattice parameter for aforementioned BA type of covering, see Fig. 5.

The electronic spectrum of $Ti_3CN(OH)_2$ phases includes already four subbands, composed of hybridized and merged O2*p*-Ti3*d* and N2*p*-Ti3*d* states, the separated subband of hybridized C2*p*-Ti3*d* states, and the near-Fermi subband of Ti3*d* states, Fig.

7. We found that the hydroxylated $Ti_3CN$ retain the metallic-like behavior (Figs. 6 and 7) – unlike some carbon-containing *MX*enes, which upon functionalization by hydroxy groups can become semiconductors [18,21]. The reason is enough obvious and in framework of rigid electronic band model is explained by the increased electron count in canbonitride *MX*enes from nitrogen – as compared with "pure" carbon-containing MXenes.

*3.3. The influence of distribution of carbon and nitrogen atoms and of the ratio C/N on some properties of hydroxylated $Ti_3C_{2-x}N_x(OH)_2$.*

It is well known that for conventional solid solutions of carbonitrides $TiC_{1-x}N_x$ with NaCl-like lattice, the C/N atomic distribution in anion sublattice has random character [1-5]. For MAX carbonitrides: $Ti_2C_{1-x}N_x$ [28,29] and $Ti_3C_{2-x}N_x$ [28] also there is no clear evidence for an ordered distribution of C and N atoms. Another remarkable feature of these materials is a broad C/N compositions range: for NaCl-like carbonitrides $TiC_{1-x}N_x$ the degree of homogeneity $x$ may vary from 0 to 1. The MAX carbonitrides have been also synthesized for a varied ratio C/N [28,29].

Therefore, on the example of hydroxylated 2D titanium carbonitrides, we have also examined the influence of distribution of carbon and nitrogen atoms and of the ratio C/N on the electronic properties and on the relative stability of these materials. In all these calculations, the AA type of the OH-covering, and the random distribution of C and N atoms were assumed. In this manner, we have considered the series of *MX*enes $Ti_3C_{2-x}N_x(OH)_2$ with the variation of the C/N ratio from $x = 0.0$ to $x = 2.0$. The results are depicted in Figs. 6-8.

Yet, the integrity of the layers is preserved, the geometry optimization reveals an internal distortion of $Ti_3C_{2-x}N_x(OH)_2$ layers and a broadened distribution of the bond lengths (Fig. 2.5). It indicates the possibility for the modification of the DOS profiles comparing to the "ideal" undistorted lattice of $Ti_3CN(OH)_2$.

In general, the electronic spectra of $Ti_3C_{2-x}N_x(OH)_2$ (Fig. 7.3-7.5) are similar to those for ordered $Ti_3CN(OH)_2$ phase (Fig. 7.2) and include four main subbands of O2*p*-Ti3*d*, N2*p*-Ti3*d*, C2*p*-Ti3*d* states, and the near-Fermi subband of Ti3*d* states. Though, the random distribution of C and N atoms leads to the smoothing and merging of separated peaks on the DOS profile. With growth of N/C relation the intensity of N2*p*- like DOS peak increases. Accordingly, the intensity of C2*p*-like DOS peak vanishes. All of studied $Ti_3C_{2-x}N_x(OH)_2$ layers with random distribution of C and N are metallic-like, but in contrast to the ordered $Ti_3CN(OH)_2$ the Fermi level is placed in the local DOS minimum, Fig. 7.

Finally, to provide an insight into the stability of studied *MX*enes $Ti_3C_{2-x}N_x(OH)_2$, their enthalpy of mixing $\Delta H$ was estimated. Assuming the formation of these carbonitrides as the solid solutions of the corresponding *MX*enes $Ti_3C_2(OH)_2$ and $Ti_3N_2(OH)_2$ in the formal reaction: $Ti_3C_{2-x}N_x(OH)_2 \rightarrow ½(2-x)Ti_3C_2(OH)_2 + ½xTi_3N_2(OH)_2$, the values of $\Delta H$ were calculated as: $\Delta H = E_{tot}(Ti_3C_{2-x}N_x(OH)_2) - ½(2-x)E_{tot}(Ti_3C_2(OH)_2) - ½xE_{tot}(Ti_3N_2(OH)_2)$, where $E_{tot}$ are the total energies of the corresponding systems as obtained in our SCC-DFTB calculations.

The results depicted in Fig. 8 show that $\Delta H < 0$, indicating that the formation of *MX*ene carbonitrides is energetically favorable even due to the enthalpy factor. Enrichment of mixed $Ti_3C_{2-x}N_x(OH)_2$ compositions with N content should facilitate their stabilization. The absolute value $\Delta H$ for composition $Ti_3CN(OH)_2$ with random distribution of C and N atoms is larger, than that for ordered $Ti_3CN(OH)_2$. Thus, formation of hydroxylated carbonitride solid solutions with random distribution of C and N atoms would be favored due to both mixing enthalpy and entropy factors.

## 4. Conclusions

In summary, in this paper we have probed the most recently synthesized nano-sized 2D titanium carbonitrides and their hydroxylated derivatives – so-called *MX*enes $Ti_3C_{1-x}N_x$ and $Ti_3C_{1-x}N_x(OH)_2$ - by means of SSC-DFTB calculations. An insight into the peculiarities of the structural, electronic properties and stability of these materials was focused on their genesis in the sequence: binary *MX*enes $Ti_3C_2$ ($Ti_3N_2$) → their hydroxilated forms $Ti_3C_2(OH)_2$ ($Ti_3N_2(OH)_2$) → pristine *MX*ene carbonitrides → their hydroxilated derivatives.

We have found that for pristine *MX*ene carbonitrides the replacement of carbon by nitrogen leads to decrease in the lattice parameters relative to binary *MX*ene $Ti_3C_2$. These materials are metallic-like, their electronic spectra include three separate subbands composed of hybridized N2*p*-Ti3*d*, of hybridized C2*p*-Ti 3*d* states forming covalent Ti-N and Ti-C bonds and near-Fermi Ti3*d* states.

For hydroxylated *MX*ene carbonitrides the properties are studied depending on (i) the type of surface termination by OH groups, (ii) type of distribution of carbon and nitrogen atoms (ordering/disordering effect) and (iii) the ratio C/N (compositions range). The results show that the lattice parameters of hydroxylated *MX*ene carbonitrides depend strongly on the type of arrangement of OH groups. The most favorable covering corresponds to one with all OH groups placed at the hollow sites between three neighboring carbon (nitrogen) atoms of carbonitride layer. These materials should be also metallic-like. Their electronic spectra include four subbands - of hybridized O2*p*-Ti3*d*, N2*p*-Ti3*d*, C2*p*-Ti3*d* states, which are responsible for Ti-O, Ti-N and Ti-C directional bonds, and the near-Fermi subband of Ti3*d* states. Besides, our results reveal that the formation of hydroxylated carbonitrides with random distribution of C and N atoms is favorable like for the related 3D systems: NaCl-like solid solutions - carbonitrides $TiC_{1-x}N_x$, as well as for MAX carbonitrides such as $Ti_3C_{2-x}N_x$ or $Ti_2C_{1-x}N_x$, and the most stable compositions of $Ti_3C_{2-x}N_x(OH)_2$ should be nitrogen-rich ($x > 1$).

Finally, in this paper we focused on the “stoichiometric” (Ti/(C+N)=3/2) 2D titanium carbonitrides $Ti_3C_{2-x}N_x$ and their hydroxylated derivatives $Ti_3C_{2-x}N_x(OH)_2$. We believe that the further prospects for tailoring of the functional properties of these interesting 2D materials can be achieved at least in three ways: (i) by variation of the amount of the vacancies in (C,N) sublattice (about the presence of the considerable amount of the lattice vacancies in parent MAX phases see, for example, Refs. [28,29]); (ii) by covering of 2D *MX*ene carbonitrides by various adatoms or functional groups, or (iii) by modulation of the degree of surface termination by these adatoms or functional groups. The further experimental and theoretical studies seem very desirable to understand these effects, which can promote the possible applications of these new 2D materials.

* * *

**Table 1**. The relative energies (ΔE, eV/atom) for *MX*enes $Ti_3C_2$, $Ti_3N_2$, and $Ti_3CN$, functionalized by hydroxy groups as obtained within SSC-DFTB calculations.

| structure* | Δ*E* | structure* | Δ*E* |
|---|---|---|---|
| $Ti_3C_2(OH)_2$ (AA) | 0.0000 | | |
| $Ti_3C_2(OH)_2$ (BB) | 0.0717 | $Ti_3CN(OH)_2$ (AA) | 0.0000 |
| $Ti_3C_2(OH)_2$ (AB) | 0.0300 | $Ti_3CN(OH)_2$ (BB) | 0.0964 |
| $Ti_3N_2(OH)_2$ (AA) | 0.0000 | $Ti_3CN(OH)_2$ (AB) | 0.0875 |
| $Ti_3N_2(OH)_2$ (BB) | 0.0760 | $Ti_3CN(OH)_2$ (BA) | 0.0299 |
| $Ti_3N_2(OH)_2$ (AB) | 0.0654 | | |

* the types of OH covering (AA, AB *etc*) see Fig. 1

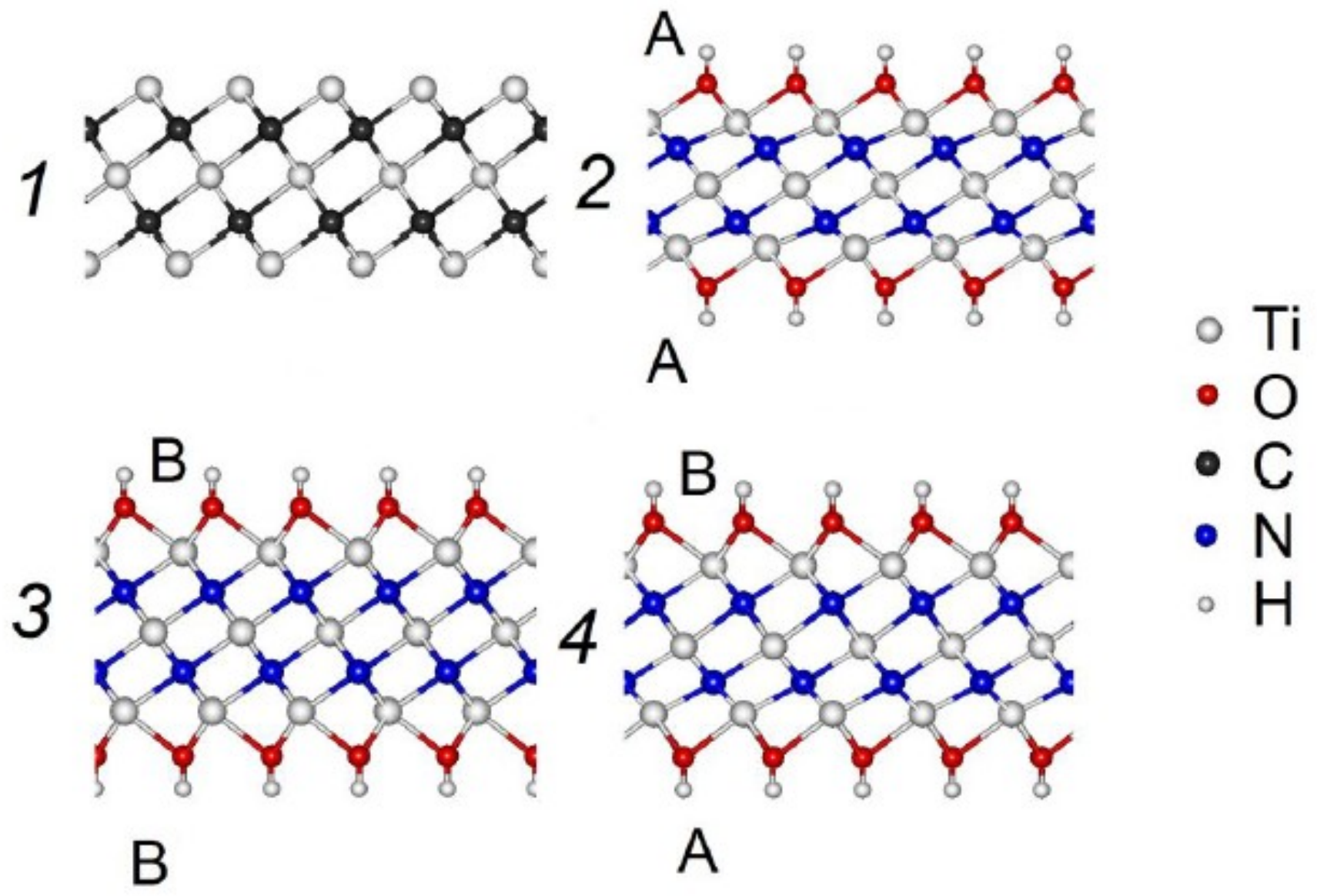

**Fig. 1**. The side views of optimized atomic structures of free-standing *MXene* $Ti_3C_2$ (*1*) and of the hydroxylated derivatives of *MXene* $Ti_3N_2$ - $Ti_3N_2(OH)_2$ with possible configurations of OH covering on external Ti sheets:  A – all OH groups are placed at the hollow site between three neighboring N atoms, and B – all OH groups are placed at the top site of N atom. For the double-sided surface termination, three main configurations of $Ti_3N_2(OH)_2$ are depicted: two symmetric AA (*2*) and BB (*3*), and the asymmetric configuration AB (*4*).

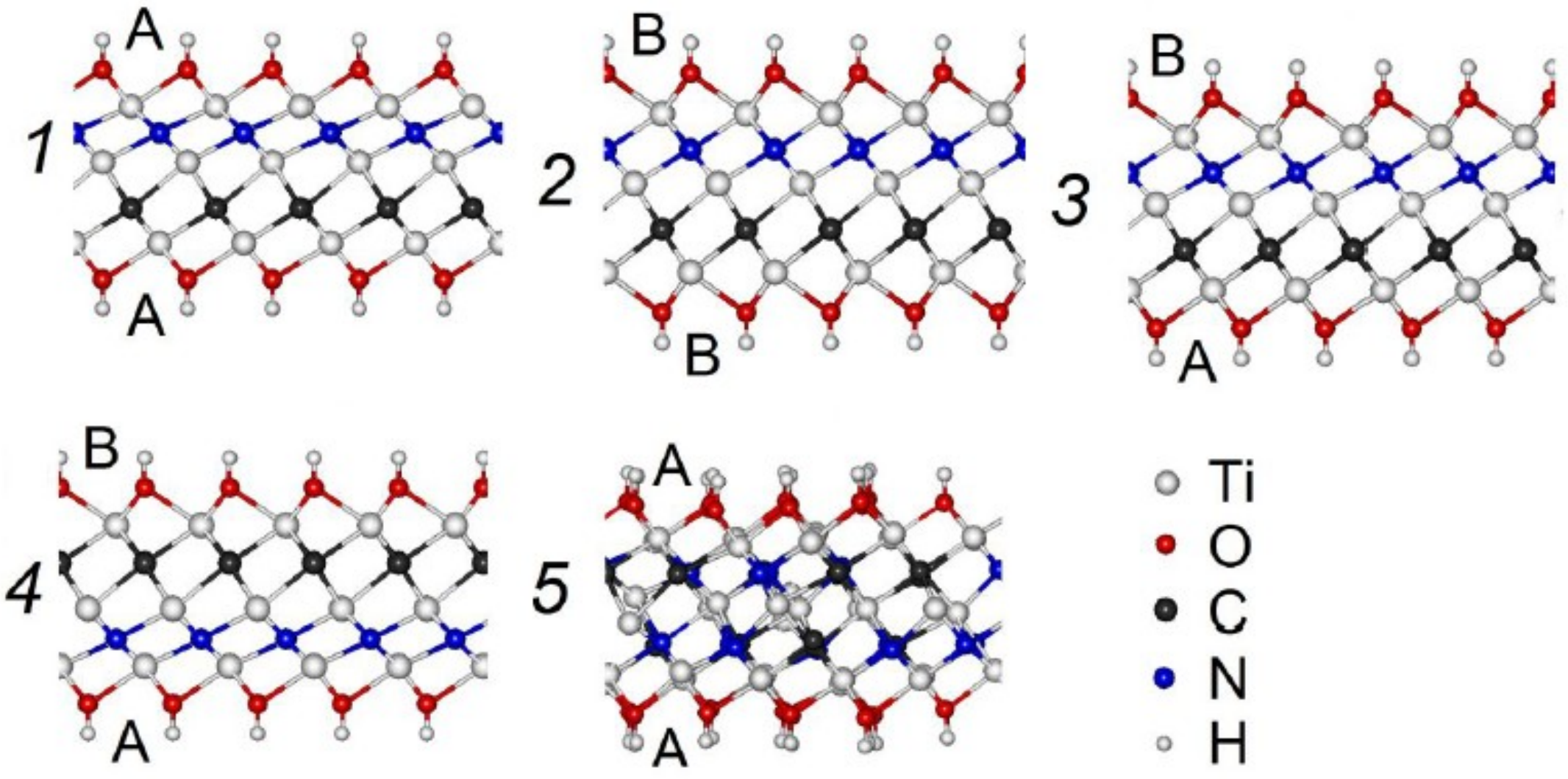

**Fig. 2.** The side views of optimized atomic structures of hydroxylated derivatives of *MXene* $Ti_3CN$ with the ordered configuration, where carbon and nitrogen atoms form the individual atomic sheets (*1-4*) within the layer, and with various types of surface termination: AA (*1*), BB (*2*), AB (*3*), and BA (*4*). Structure (5) represents an example of random distribution of C and N atoms  and with AA type of the surface termination by hydroxy groups.

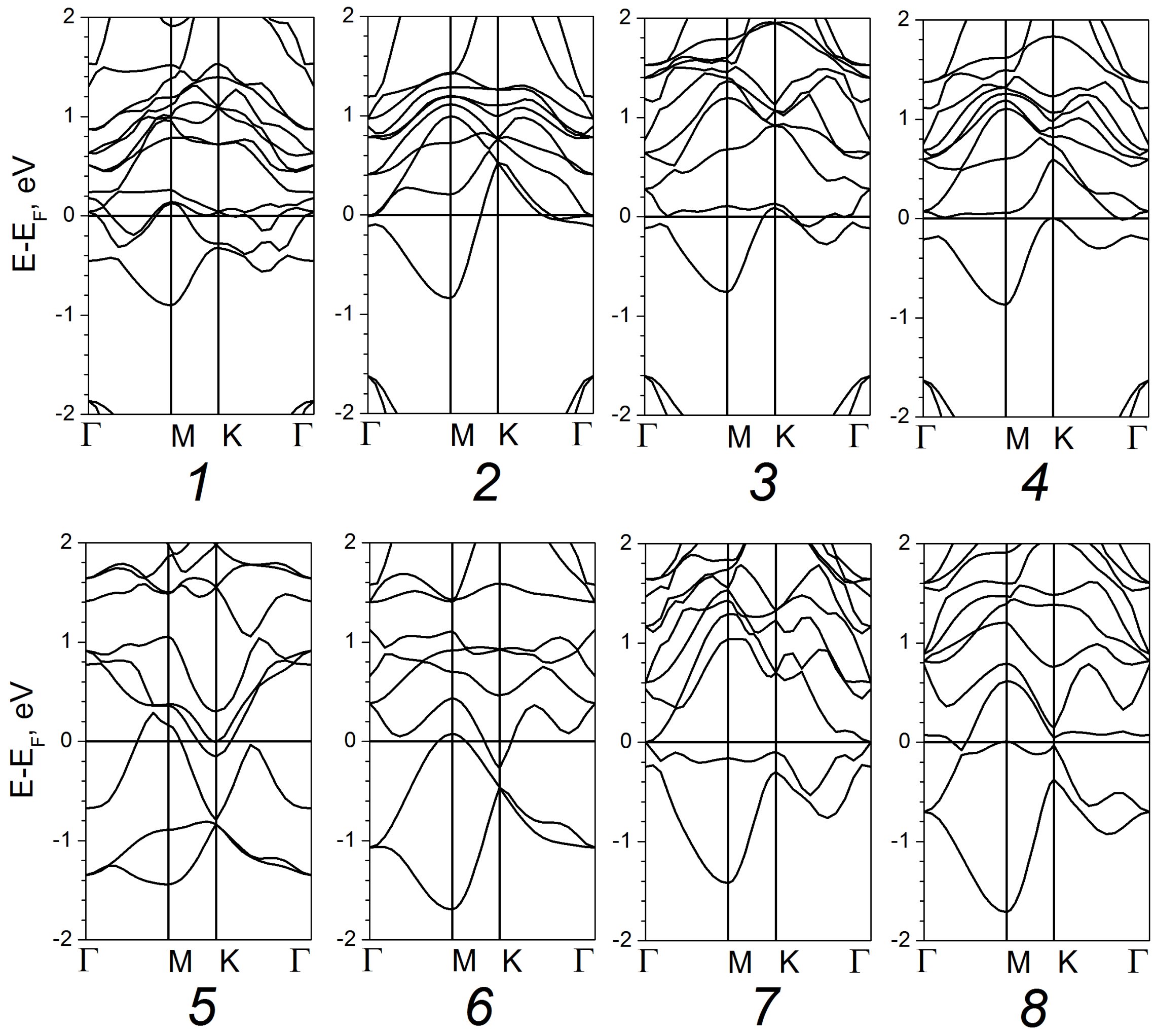

**Fig. 3.** *Top panel*: near-Fermi electronic bands for bare *MXene* $Ti_3C_2$ (*1*) and for its hydroxylated derivatives $Ti_3C_2(OH)_2$ (*2-4*) with possible configurations of OH covering of external Ti sheets: AA (*2*), BB (*3*), AB (*4*). *Bottom panel*: near-Fermi electronic bands for bare *MXene* $Ti_3N_2$ (*5*) and for hydroxylated derivatives $Ti_3N_2(OH)_2$ (*6-8*) with possible configurations of OH covering: AA (*6*), BB (*7*), AB (*8*). The types of OH configurations see Fig. 1.

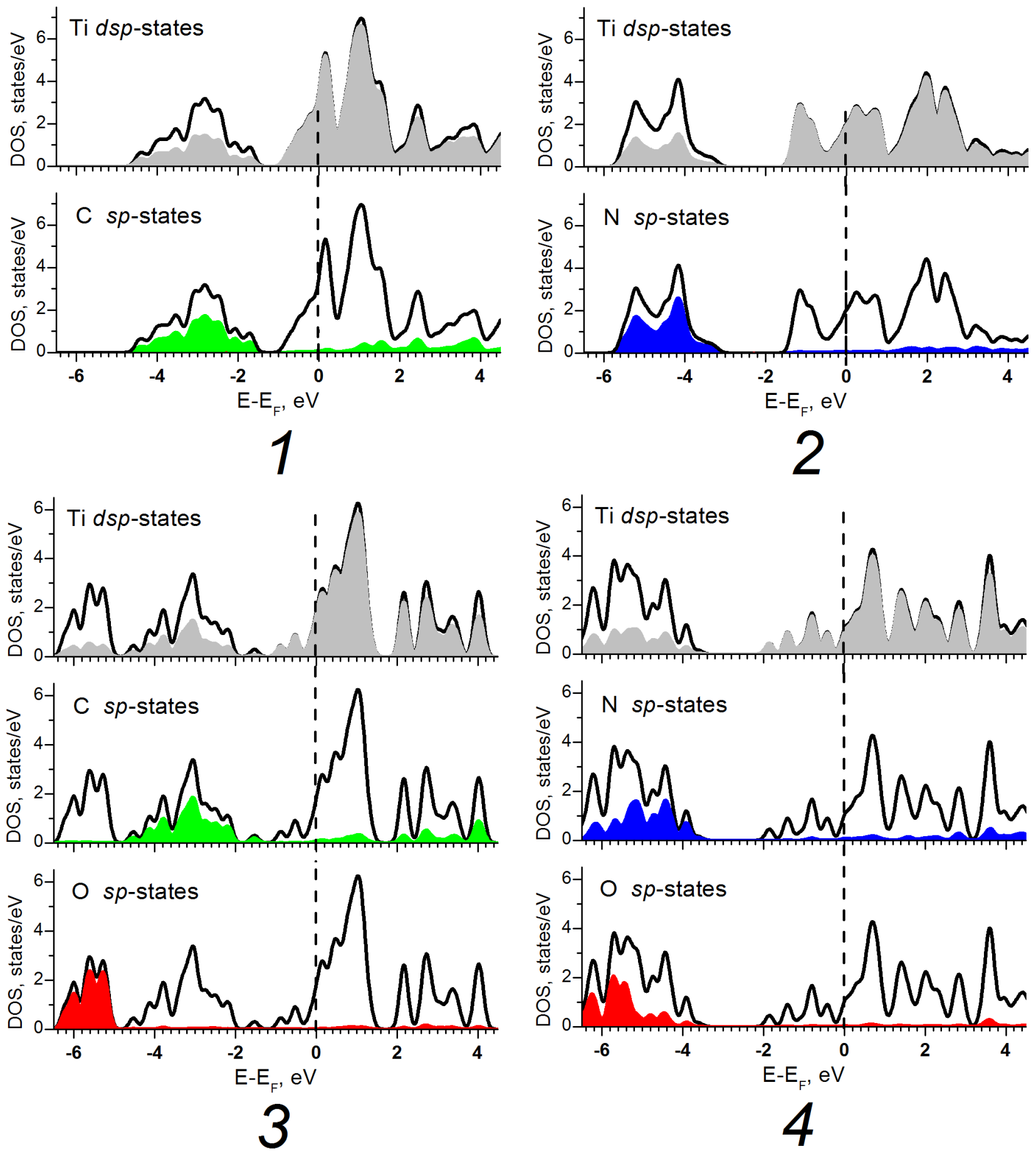

**Fig. 4.** Total and atomic-projected densities of states (DOSs) for bare *MXenes* $Ti_3C_2$ (*1*), and $Ti_3N_2$ (*2*) and their most stable hydroxylated derivatives $Ti_3C_2(OH)_2$ (*3*) and $Ti_3N_2(OH)_2$ (*4*) with AA arrangement of OH-groups.

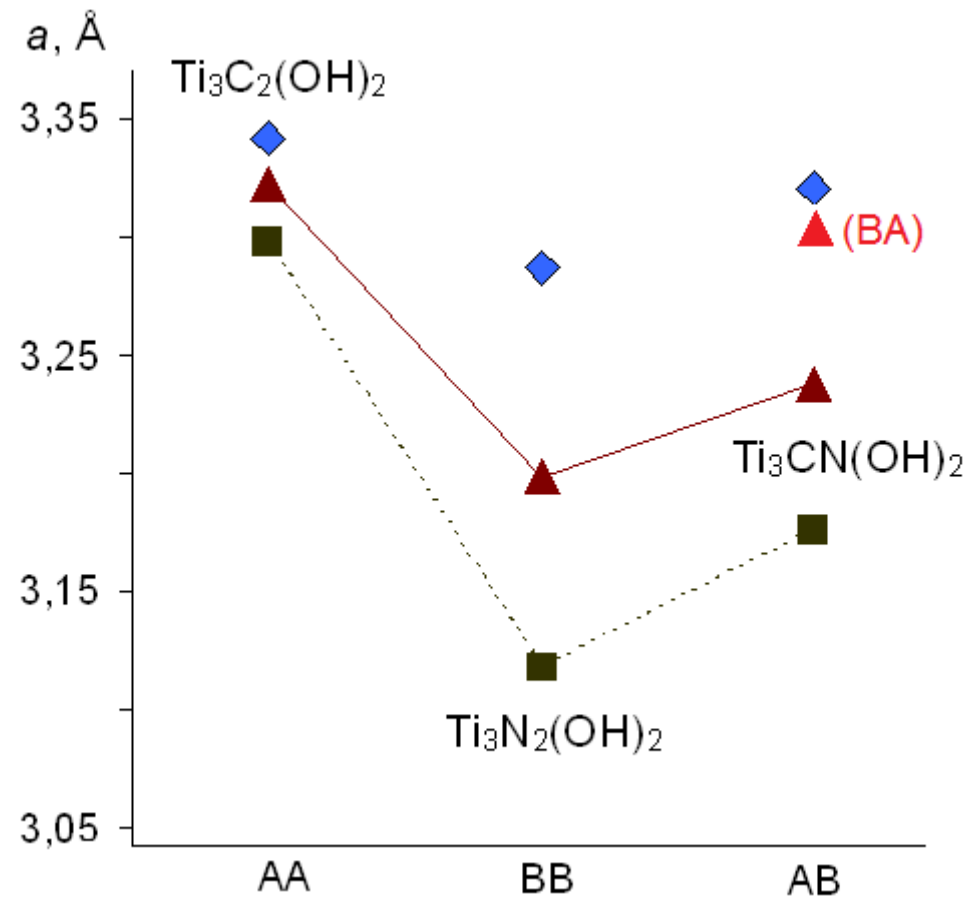

**Fig. 5.** Calculated lattice parameters for hydroxylated MXenes $Ti_3C_2(OH)_2$, $Ti_3N_2(OH)_2$, and $Ti_3CN(OH)_2$ with the covering types AA, BB, and AB. For $Ti_3CN(OH)_2$, the lattice parameter for additional possible covering type (BA, see text) is given.

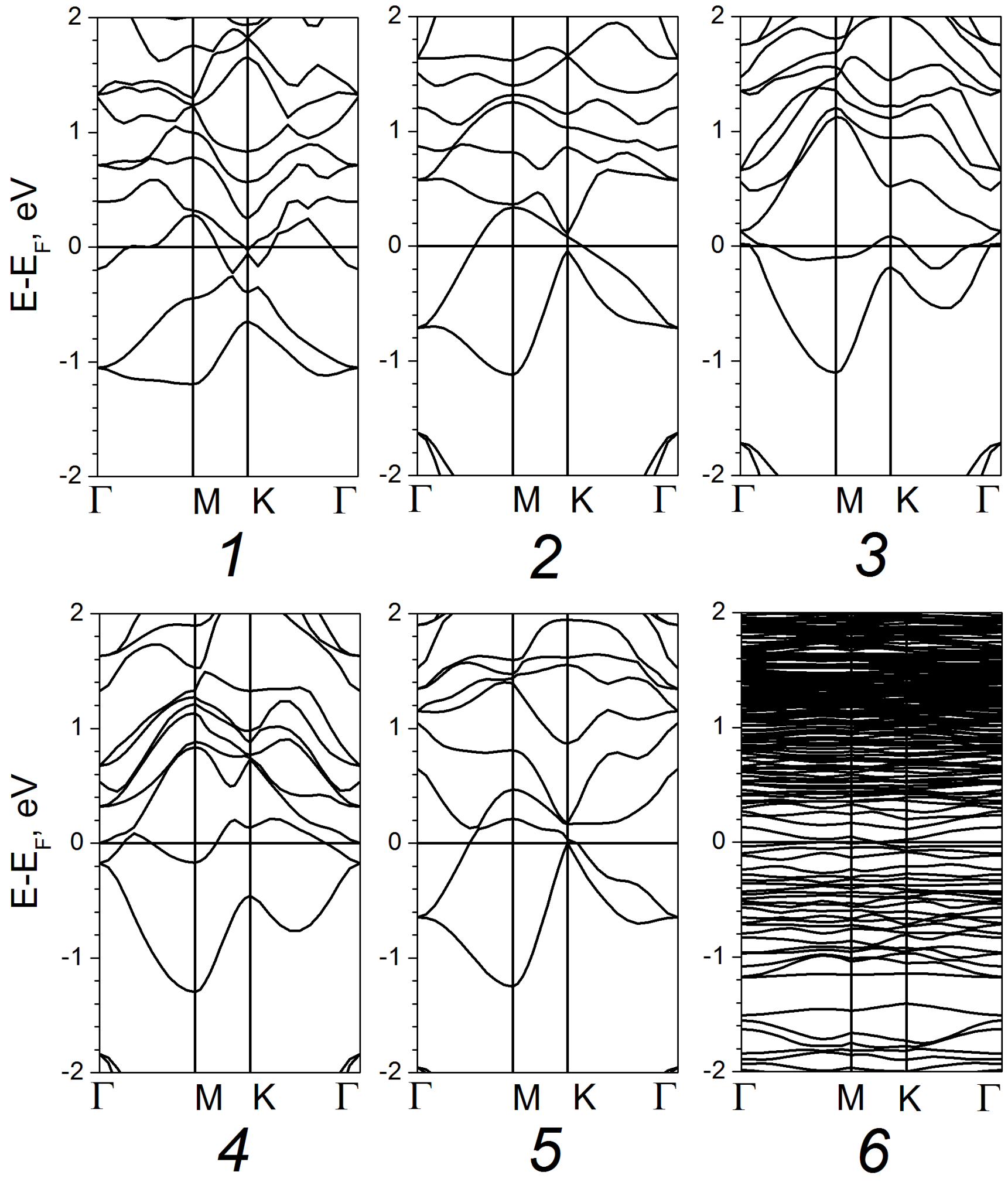

**Fig. 6.** Near-Fermi electronic bands for ordered *MXene* $Ti_3CN$ (*1*) and for its hydroxylated derivatives $Ti_3CN(OH)_2$ (*2*-*5*) with various types of OH covering: AA (*2*), BB (*3*), AB (*4*), BA (*5*). The electronic bands of disordered $Ti_3CN(OH)_2$ with covering type AA (*6*).

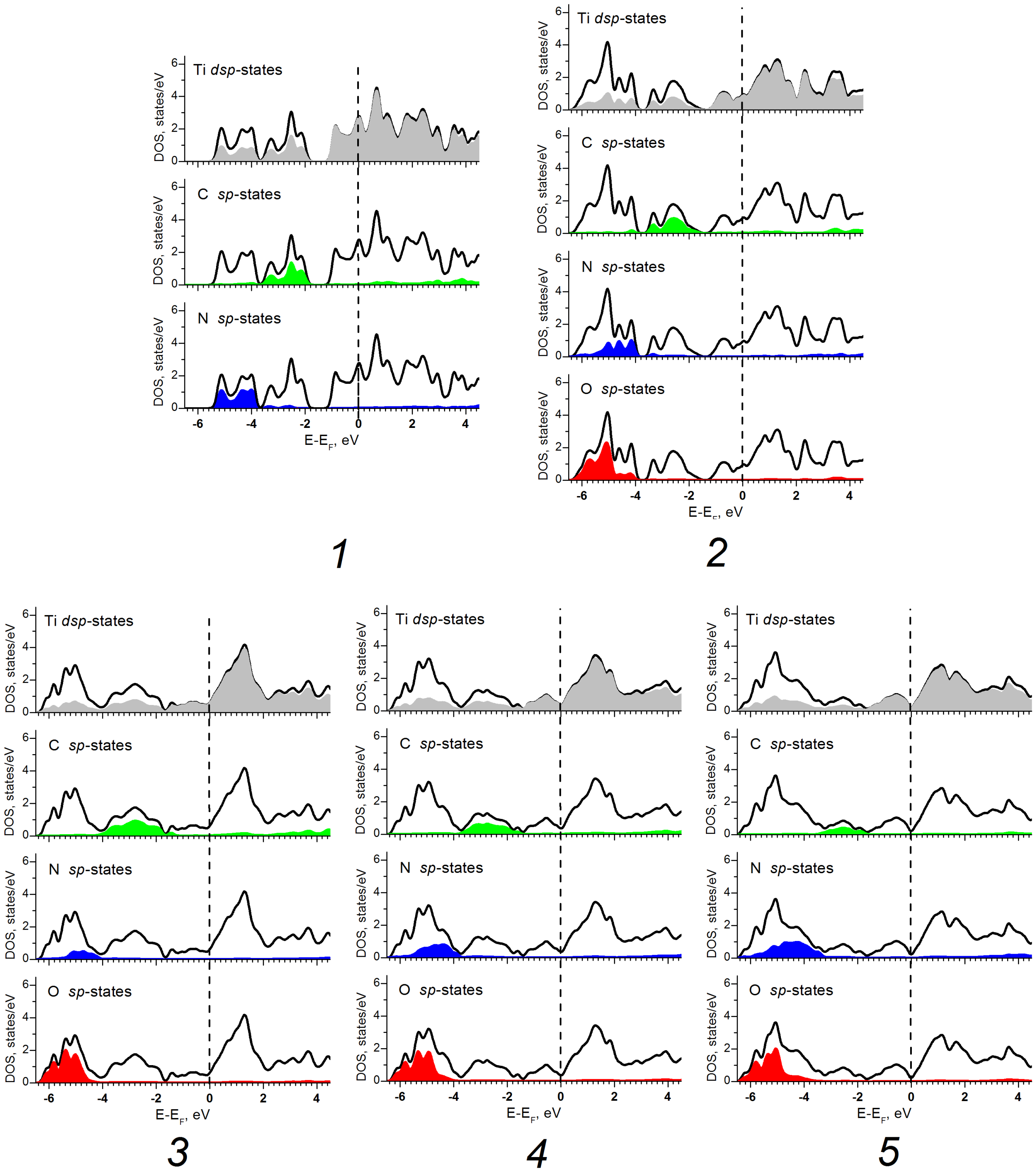

**Fig. 7.** Total and atomic-projected densities of states (DOSs) for ordered *MX*ene $Ti_3CN$ (*1*) and for its most stable hydroxylated derivative $Ti_3CN(OH)_2$ with AA-type of OH covering (*2*), as well as for the hydroxylated derivatives of *MX*enes $Ti_3C_{2-x}N_x$ with various C/N ratio and with random distribution of C and N atoms: $Ti_3C_{1.5}N_{0.5}(OH)_2$ (*3*), $Ti_3CN(OH)_2$ (*4*), $Ti_3C_{0.5}N_{1.5}(OH)_2$ (*5*).

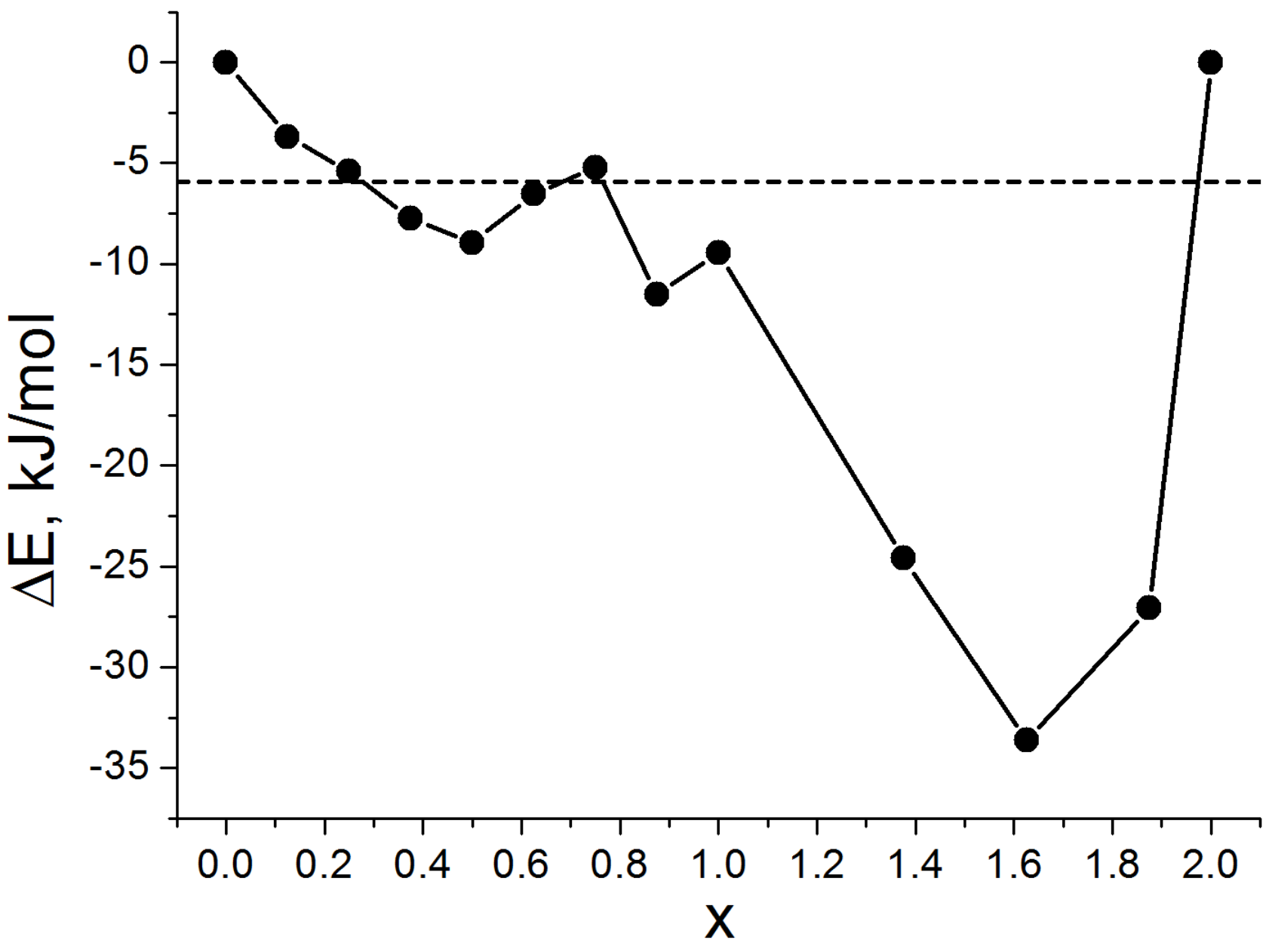

**Fig. 8.** Calculated enthalpy of mixing for the formation of solid solutions $Ti_3C_{2-x}N_x(OH)_2$ with random distribution of C and N atoms (Fig. 2.5). The enthalpy for composition $Ti_3CN(OH)_2$ with ordered uniplanar distributions of C and N atoms (Fig. 2.1) is given as the dashed line.